# MgO thickness-induced spin reorientation transition in $Co_{0.9}Fe_{0.1}/MgO/Co_{0.9}Fe_{0.1}$ structure.

K. Lasek, L. Gladczuk, M. Sawicki, P. Aleshkevych and P. Przyslupski

Institute of Physics, Polish Academy of Sciences, Aleja Lotnikow 32/46, PL-02668 Warsaw, Poland

Email: klasek@ifpan.edu.pl

## Abstract

The magnetic anisotropy (MA) of $Mo/Au/Co_{0.9}Fe_{0.1}/Au/MgO(0.7 - 3\ nm)/Au/Co_{0.9}Fe_{0.1}/Au$ heterostructure has been investigated at room temperature as a function of MgO layer thickness ($t_{MgO}$). Our studies show that while the MA of the top layer does not change its character upon variation of $t_{MgO}$, the uniaxial out-of-plane MA of the bottom one undergoes a spin reorientation transition at $t_{MgO}$ of about 0.8 nm, switching to the regime where the coexistence of in- and out-of-plane magnetization alignments is observed. The magnitudes of the magnetic anisotropy constants have been determined from ferromagnetic resonance and dc-magnetometry measurements. The origin of MA evolution has been attributed to a presence of an interlayer exchange coupling (IEC) between $Co_{0.9}Fe_{0.1}$ layers through the thin MgO film.

## 1. Introduction

Ferromagnetic-metal/insulator/ferromagnetic-metal (FM/I/FM) magnetic tunneling junctions (MTJs) are fundamental building elements for the realization of contemporary spintronic devices such as spin-transfer-torque magnetic random access memory devices STT-MRAM [1] or nanoscales oscillators with enormous output power.[2] Typical MTJs are composed of electrodes with an in-plane magnetization.[3,4] However, MTJs with perpendicular magnetic anisotropy (p-MTJs) have great advantages over the in-plane ones, due to their high tunnel magnetoresistance ratio (TMR), high thermal stability and low critical current for current induced magnetization switching.[5-7] Therefore, the experimental determination and understanding of magnetic anisotropy and mechanisms of its control in ferromagnetic thin films is crucial towards the MTJ system design for future use in electronic applications. As the thickness of a magnetic layer is reduced to few monolayers, the magnetic anisotropy energy (MAE) in addition to its magnetocrystalline part acquires important contributions from the surface and the oblate shape of the material. The change of the balance between these quantities can result in the reorientation of the anisotropy axis - the spin reorientation transition (SRT). This phenomenon has been observed either as a function of temperature,[8-11] or thickness of the: magnetic layer [10-14] or capping layer.[15,16] In addition, oscillations of the magnetic anisotropy (MA) as a function of the MgO barrier thickness in the MTJs composed of the binary multilayer Co/Pt and Co/Pd were observed.[17] These anisotropy variations were assigned to the film morphology. However, no systematic studies of an influence of the MgO thickness on the MA, especially near the SRT, in FM/I/FM structures have been available so far.

Here we report the results of the investigations of the MgO spacer thickness effect on the magnetic anisotropy in $Co_{0.9}Fe_{0.1}$/Au/MgO($t_{MgO}$)/Au/$Co_{0.9}Fe_{0.1}$ MTJ structure in the low $t_{MgO}$ regime, $1.0 > t_{MgO} > 0.7$ nm. Experiments, accomplished by the ferromagnetic resonance spectroscopy (FMR) and superconducting quantum interference device magnetometry (SQUID), have proven that on entering this range of $t_{MgO}$, the uniaxial MA of one of the constituent $Co_{0.9}Fe_{0.1}$ layers consistently changes from the out-of-plane to the easy-plane one. The corresponding quenching of the effective uniaxial anisotropy constant allows us to establish the magnitude of the fourth-order component to the magnetic anisotropy energy of this layer.

## 2. Experimental

The studied heterostructures have been deposited by molecular beam epitaxy (MBE) method onto a–plane sapphire substrates. The growth of each structure has been seeded with Mo(110) and continued with 20 nm thick Au(111) buffer layers. The $Co_{0.9}Fe_{0.1}$ alloy layers of thickness d = 1.4 nm, the main components of the investigated structures, have been obtained by a co-deposition of Co and Fe using an electron gun and an effusion cell, respectively. In order to improve perpendicular magnetic anisotropy and minimize effects of interface mixing or possible diffusion in a $Co_{0.9}Fe_{0.1}$ layer, the MgO spacer of each $Co_{0.9}Fe_{0.1}$ layer has been preceded and covered with a monoatomic layer of Au.[18] A special approach was necessary to assure equivalence of the FM electrodes in specimens with different $t_{MgO}$. To this end, a master MTJ sample was prepared in a shape of about 1 cm long strip in which a wedge of MgO spacer has been grown. Enumerating from the substrate, the whole stack is composed of Mo/Au/$Co_{0.9}Fe_{0.1}$/Au/MgO-wedge/Au/$Co_{0.9}Fe_{0.1}$/Au. The MgO wedge has been accomplished by moving a shutter at a constant speed of 0.05 mm/s close to the sample surface during the MgO deposition. In such obtained sample $t_{MgO}$ varied from 0 to 1 nm with a slope of 0.1 nm/mm. After completion of the growth, the sample has been cleaved perpendicularly to the wedge direction into pieces of approximately 1 mm wide. For the study reported here, we chose a representative group of samples with $t_{MgO}$ of 0.7, 0.78, 0.85, 0.92, and 1.0 nm denoted collectively as $S_{MTJ}$. To independently verify the magnetic properties of the bottom ($F_B$) and the top ($F_T$) $Co_{0.9}Fe_{0.1}$ electrodes of the $S_{MTJ}$ stack two main references have been prepared: Au/$Co_{0.9}Fe_{0.1}$/Au/MgO-wedge/Au and Au/MgO/Au/$Co_{0.9}Fe_{0.1}$/Au - an $R_B$ and $R_T$ samples, respectively. Additionally, for evaluation of MA of the FM/I/FM structure with well-separated FM electrodes another sample, Au/$Co_{0.9}Fe_{0.1}$/Au/MgO( 3 nm)/Au/$Co_{0.9}Fe_{0.1}$/Au with MgO thickness of 3 nm has been fabricated (denoted as $R_{MTJ}$). After completing the deposition, all the samples have been covered with 10 nm thick Au capping layer to protect them from oxidation during the ex-situ measurements. Table 1. summarizes all the investigated samples in this study.

Table 1. The active part of the structure of all the investigated samples. $F_B$ ($F_T$) denotes the ferromagnetic $Co_{0.9}Fe_{0.1}$ layer deposited before (after) the deposition of the MgO layer of a thickness $t_{MgO}$.

| Sample ID | Structure | $t_{MgO}$ (nm) |
|---|---|---|
| $S_{MTJ}$ | Au/$F_B$/Au/MgO($t_{MgO}$)/Au/$F_T$/Au | 0.7 – 1.0 |
| $R_B$ | Au/$F_B$/Au/MgO($t_{MgO}$)/Au | 0.5 – 1.2 |
| $R_T$ | Au/MgO($t_{MgO}$)/Au/$F_T$/Au | 1.5 |
| $R_{MTJ}$ | Au/$F_B$/Au/MgO($t_{MgO}$)/Au/$F_T$/Au | 3.0 |

The ferromagnetic resonance (FMR) experiments were carried out using Bruker EMX spectrometer working at a fixed frequency $f$ = 9.38 GHz. The magnetic hysteresis $M(H)$ loops were measured with an MPMS XL Quantum Design SQUID magnetometer.[19] All the experiments reported here have been performed at room temperature.

## 3. Results

### 3.1. Ferromagnetic resonance

It turns out that the magnetic anisotropy of the $Co_{0.9}Fe_{0.1}$ layer depends on the position of the layer in the MTJ stack. This notion comes from a comparison of the ferromagnetic resonance spectra of the reference $R_B$, $R_T$, and $R_{MTJ}$ samples, presented in the inset of Fig. 1, measured with the external magnetic field ($H_{ext}$) applied perpendicularly to the surface of the samples. Since all the constituent $Au/Co_{0.9}Fe_{0.1}/Au$ stacks are nominally the same, so a larger (smaller) magnitude of $H_{res}$ observed for the $R_T$ ($R_B$) reference indicates the in-plane (perpendicular) MA, respectively. For this comparison, the spectrum of only one sample from the $R_B$ series has been selected (with $t_{MgO} = 1$ nm) since, as detailed later, the MA of $Co_{0.9}Fe_{0.1}$ layer shows only marginally weak sensitivity on the thickness of the MgO overlayer. Further, we note that the spectrum of the $R_{MTJ}$ structure, containing two similar FM layers separated by 3 nm thick MgO spacer, consists of two resonances located at similar fields as for the $R_T$ and $R_B$ references, namely around 1.5 and 8 kOe, respectively. This finding indicates that the MA of the FM layers in the MTJ structures is predominantly associated with the MgO and FM layers growing sequence and that in the MTJ stack the bottom $Co_{0.9}Fe_{0.1}$ layer exhibits perpendicular MA while the top one exhibits the in-plane MA.

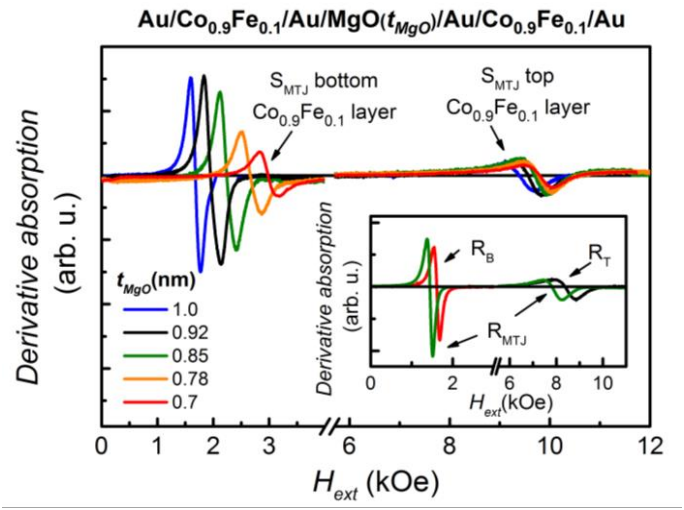

Fig. 1. (Color online). The out-of-plane ferromagnetic resonance (FMR) spectra of all the investigated $Au/Co_{0.9}Fe_{0.1}/Au/MgO(t_{MgO})/Au/Co_{0.9}Fe_{0.1}/Au$ - $S_{MTJ}$ samples. (Inset) The out-of-plane FMR spectra of the reference $Au/Co_{0.9}Fe_{0.1}/Au/MgO(1.0$ nm$)/Au$ - $R_B$, $Au/MgO(1.5$ nm$)/Au/Co_{0.9}Fe_{0.1}/Au$ - $R_T$, and $Au/Co_{0.9}Fe_{0.1}/Au/MgO(3$ nm$)/Au/Co_{0.9}Fe_{0.1}/Au$ - $R_{MTJ}$ samples.

To confirm this conjecture a dependence of the FMR resonance field on the angle between external magnetic field and the normal to the samples' surface ($\theta_H$) have been measured. Indeed, both the $H_{res}(\theta_H)$ of the $R_B$ (Fig. 2a) and $R_T$ (Fig. 2b) samples, clearly show a 180 degrees periodicity - implying a dominant uniaxial contribution, and their relative 90 degrees shift in $\theta_H$ confirms mutually perpendicular easy axis orientation: along the normal to the samples' surface (perpendicular MA) for the $F_B$ layer and the easy-plane MA for the $F_T$ layer. Importantly, the same contributions are simultaneously observed in the $H_{res}(\theta_H)$ of $R_{MTJ}$ structure (Fig. 2c) confirming that for large separations ($t_{MgO}$ = 3 nm) these two $Co_{0.9}Fe_{0.1}$ layers stay uncoupled, [20] retaining their MA determined by the growth sequence as in single $R_B$ and $R_T$ layers.

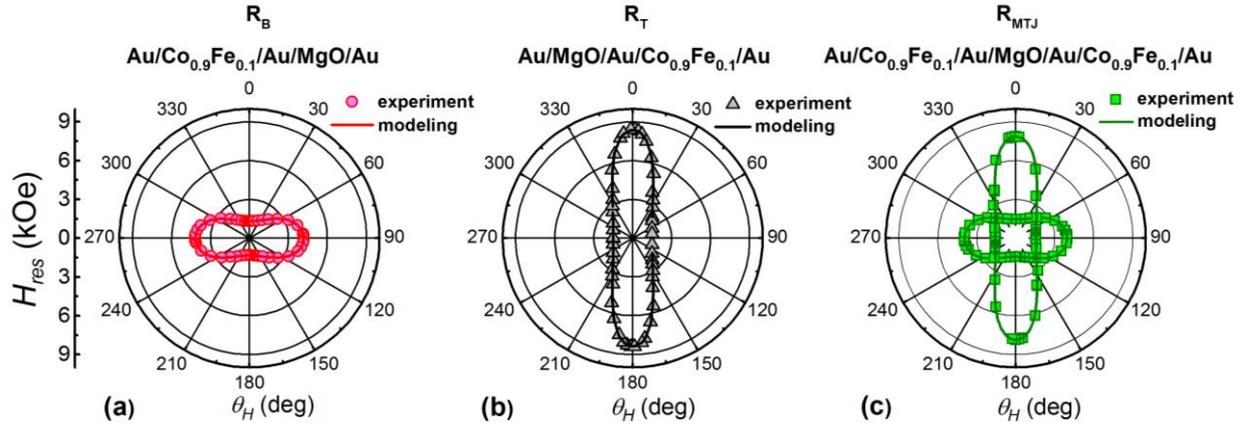

Fig. 2. Symbols: the angular dependence of the resonance field $H_{res}$ of the reference samples: (a) $R_B$, (b) $R_T$, and (c) $R_{MTJ}$. $\theta_H$ is the angle between the external magnetic field and the normal to the structure surface. The solid lines represent $H_{res}(\theta_H)$ dependences calculated according to Eq. 1 and relevant anisotropy constants given in the text.

Before turning to a detailed study of the magnetic anisotropy (MA) for MTJ system, it was crucial to select the range of the $t_{MgO}$, which is free from the discontinuities and therefore suitable for effective structural separation two $Co_{0.9}Fe_{0.1}$ layers. To this end, the FMR experiment has been conducted on the set of four $R_B$ references: $Au/Co_{0.9}Fe_{0.1}/Au/MgO$(0.5, 0.7, 1.0, and 1.2 nm)/Au, that is having fully replicated the bottom part of the MTJ stack but without the top FM layer. As it is presented in Fig. 3, for thick MgO barriers down to about 0.7 nm, the change in the anisotropy value is relatively small, indicating good barrier quality. Below this thickness, an enhancement of the perpendicular anisotropy is observed, which we attribute to the fragmentation of the MgO layer separating two gold layers which in turn increases the area of the $Co_{0.9}Fe_{0.1}/Au$ interface, characterized by greater surface energy compared to the $Co_{0.9}Fe_{0.1}/MgO$ one.[18] It is therefore concluded that down to $t_{MgO} \cong 0.7$ nm the thickness of the MgO layer itself does not play a significant role in the anisotropy evolution of the magnetic electrodes.

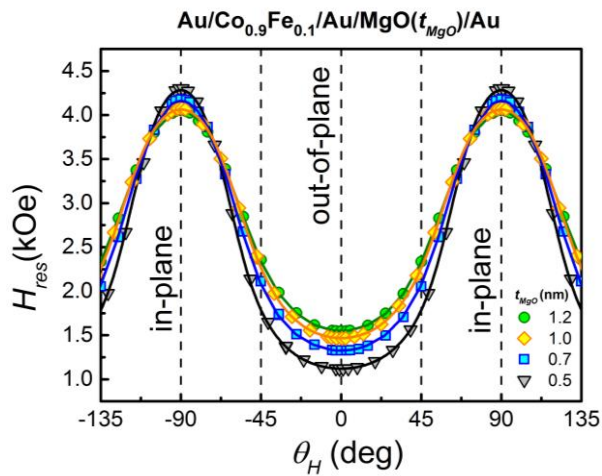

Fig. 3. The angular dependence of the resonance field $H_{res}$, of the $R_B$ - $Au/Co_{0.9}Fe_{0.1}/Au/MgO$(0.5, 0.7, 1.0, and 1.2 nm)/Au reference set of samples with MgO thickness: 0.5, 0.7, 1.0 and 1.2 nm. The solid lines represent $H_{res}(\theta_H)$ dependences calculated according to Eq. 1 and relevant anisotropy constants given in the text.

Having established the magnetic anisotropy of the $Co_{0.9}Fe_{0.1}$ layers with respect to their position in the MTJ stack, and the MgO layer thickness range suitable for further investigation, we turn to a detailed study of the MA for a MTJ system with the width of the MgO spacer reduced to 1 nm and below, in the $S_{MTJ}$ sample series. The evolution of the FMR spectra as a function of the $t_{MgO}$ in these samples is presented in the main panel of Fig. 1. It is evident that whereas the resonance of the $F_T$ layer does not change much upon reduction of the $t_{MgO}$, the resonance of the $F_B$ gets significantly shifted to larger magnetic fields. Interestingly, the magnitude of $H_{res}$ for the bottom FM layer in the $S_{MTJ}$ with $t_{MgO} = 1$ nm is nearly identical to that of the $R_{MTJ}$ reference ($t_{MgO} = 3$ nm), meaning that both FM electrodes stay fairly uncoupled even if their separation is as low as 1.6 nm, that is if we take the thickness of the very thin Au cover layers into account.

To obtain more detailed information on the influence of MgO thickness on magnetic anisotropy we have measured the out-of-plane angular dependence of resonance fields of the FMR spectra. The $H_{res}(\theta_H)$ of the bottom and the top $Co_{0.9}Fe_{0.1}$ layers of the $S_{MTJ}$ samples are shown in Fig. 4. This general overview confirms that the MA of the $F_T$ layer stays fairly unchanged exhibiting the same angular dependence and magnitude for all $t_{MgO}$. We conclude therefore that the top $F_T$ layer exhibit negligible sensitivity to the thickness of the underlying MgO spacer. Remarkable, particularly in relation to the behavior of the $F_T$, the $F_B$ layer considerable changes its MA on lowering $t_{MgO}$, as indicated in panels (b) and (c) of Fig. 4. First of all, upon lowering $t_{MgO}$, the $H_{res}(\theta_H)$ dependency changes its symmetry: from the twofold for larger $t_{MgO}$ to a fourfold-like for the lowest $t_{MgO}$ [cf. panel (b)]. This indicates a quenching of the perpendicular MA in the $F_B$ layer for substantially low $t_{MgO}$. However, on a closer inspection of Fig. 4(b) one notices that actually on going from $t_{MgO} = 1$ nm down to 0.7 nm the elongation of the $H_{res}(\theta_H)$ dependency changes from the horizontal to a vertical one. This indicates that the $F_B$ layer undergoes a SRT from the easy axis MA for $t_{MgO} \geq 0.85$ nm to the easy-plane one for $t_{MgO} \leq 0.78$ nm and that while this SRT takes place the less significant higher order anisotropy term takes a dominant role.

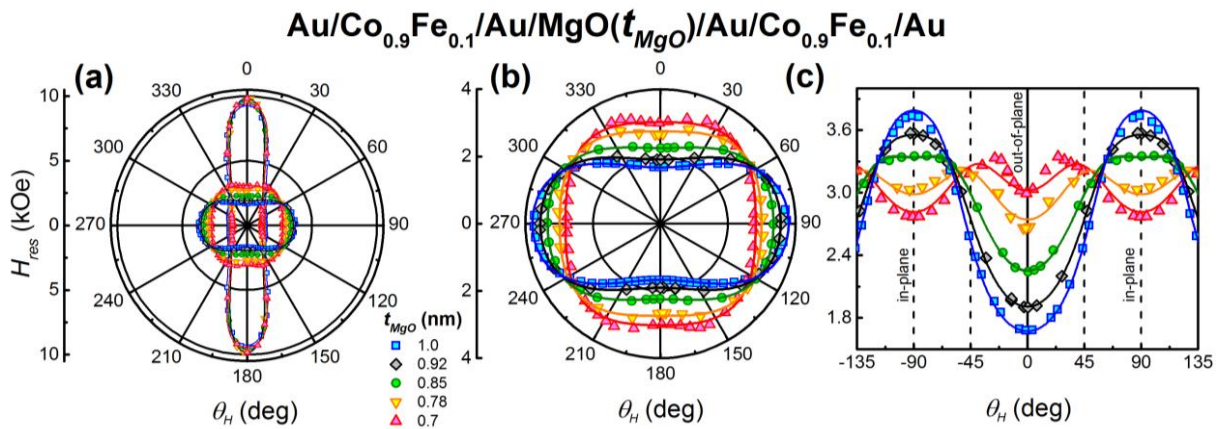

Fig. 4. Out-of-plane angular dependence of the resonance field $H_{res}$, of the $S_{MTJ}$ structures (a) a full range, (b) and (c) an expanded scales, $\theta_H$ is the angle between the external magnetic field and the normal to the structure surface. The solid lines represent $H_{res}(\theta_H)$ dependences calculated according to Eq. 1 and relevant anisotropy constants given in the text.

## 3.2 Magnetometry

Exactly the same conclusions can be drawn from our magnetometric studies performed on some of the $S_{MTJ}$ samples, as exemplified in Fig. 5. In this case a separation of the contributions specific to each magnetic layers is not possible, they contribute to the total response seen by the magnetometer in a way specific to their magnetic anisotropies. Nevertheless, since their contributions are additive, both qualitative and quantitative analysis can be performed. Firstly, one notes that the sample with $t_{MgO} = 1$ nm exhibits a two-step character, corresponding to an easy axis and a hard axis magnetization processes. This notion is substantiated by recalling that both $F_B$ and $F_T$ layers are of similar thickness and indeed, the sloppy part of the $M(H)$ starts from $M \cong M_s / 2$, where $M_s$ is the saturation magnetization of the whole stack, achieved at $H_{ext} \sim 8$ kOe. Secondly, on lowering $t_{MgO}$ the system is losing the initial easy-axis behavior, the whole $M(H)$ acquires the sloppy character for the lowest $t_{MgO}$, indicating that the MA of $F_B$ also attains the hard-axis character. Since all these measurements were performed in the perpendicular orientation of $H_{ext}$, we conclude that the mixed anisotropy case present for larger $t_{MgO}$ transforms to a case where both FM layers exhibit the easy-plane magnetic anisotropy.

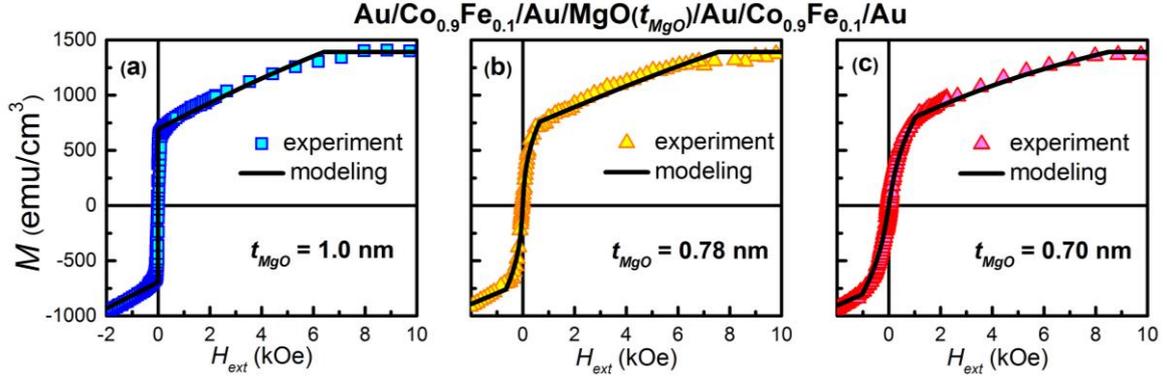

Fig. 5. Symbols: the out-of-plane magnetic hysteresis of the selected $S_{MTJ}$ samples with MgO layer thicknesses, $t_{MgO}$, equal to (a) 1.0 nm, (b) 0.78 nm, and (c) 0.70 nm. The data are corrected for the diamagnetic signal of the sapphire substrate. The solid lines represent $M(H)$ curves calculated according to Eq. 1 while the $t_{MgO}$ dependent relevant anisotropy constants are given in the text.

## 4. Modeling

Accumulated experimental evidence indicate that the dominant contribution to the magnetic anisotropy in both FM layers in the $S_{MTJ}$ stacks is the uniaxial one, however with a notable higher order contribution. The existence of the latter one is particularly evident for $t_{MgO} \leq 0.8$ nm, that is where the uniaxial component of the bottom layer undergoes the SRT from the perpendicular to the easy plane one. Following these findings we approximate the magnetostatic energy of each FM layers by [21]:

$$E = -M_s H_{ext} \cos(\theta - \theta_H) - K_{eff} \cos^2\theta - K_4 \cos^4\theta, \qquad (1)$$

The first term is the magnetic potential energy of the magnetic layer with $M$ oriented at an angle $\theta$ from its normal as the result of the application of $H_{ext}$ at the angle $\theta_H$ with respect to the normal to the layer's surface. $K_{eff}$ is the density of the effective uniaxial MA energy, which includes volume, surface, and shape anisotropies. We take $M_S = 1400 \pm 70$ emu/cm$^3$ as obtained from SQUID measurements for $H_{ext} > 8$ kOe. In the sign convention adopted here, a positive $K_{eff}$ favors the perpendicular MA. Upon the FMR data obtained for $t_{MgO} \leq 0.8$ nm, the four-fold symmetry has been described by the second-order uniaxial term (parametrized by $K_4$).

The magnitudes of $K_{eff}$ and $K_4$ are established by *simultaneous* fitting of the angular dependence of the $H_{res}$ in the FMR experiments and the $H_{ext}$-dependent SQUID magnetization results for both in- and out-of-plane ($\theta_H = 0$ and 90 deg, respectively). The condition to meet by the FMR data is obtained by evaluating the standard equation for the resonant condition given by L. Baselgia *et al.* [22] at the equilibrium position of $M_S$ ($\partial E/\partial \theta = 0$ and $\partial E/\partial \phi$). The expression for the magnetometry results is established from $\partial E/\partial\theta = 0$ and $\partial^2 E/\partial\theta^2 > 0$. Results of this procedure are shown in Figs. 2-4 and 5 as solid lines, indicating a correctness of the adopted model as the derived upon Eq. 1 conditions describe both the angular and field dependences of the magnetization very well.

The established this way dependences of the anisotropy constants magnitude on $t_{MgO}$ of all the structures researched in this study are presented in Figs. 6 (a-c). As one can see, for the thickest spacer in the $S_{MTJ}$ sample, the $F_B$ and $F_T$ electrodes [Figs. 6(a) and 6(b)] show similar MA as in the reference $R_B$, $R_T$, and $R_{MTJ}$ samples. On reducing $t_{MgO}$ $K_{eff}$ of $F_B$ changes sign crossing zero at about $t_{MgO} = 0.8$ nm, while $F_T$ only slightly changes its magnitude while keeping the sign unchanged. The $K_4$ term of $F_B$ layer [Fig. 6(c)] is an order of magnitude weaker than $K_{eff}$, and increases noticeably with lowering $t_{MgO}$. The $K_4$ for the $F_T$ layer is of a very similar strength exhibiting no systematic dependency on $t_{MgO}$ (not shown).

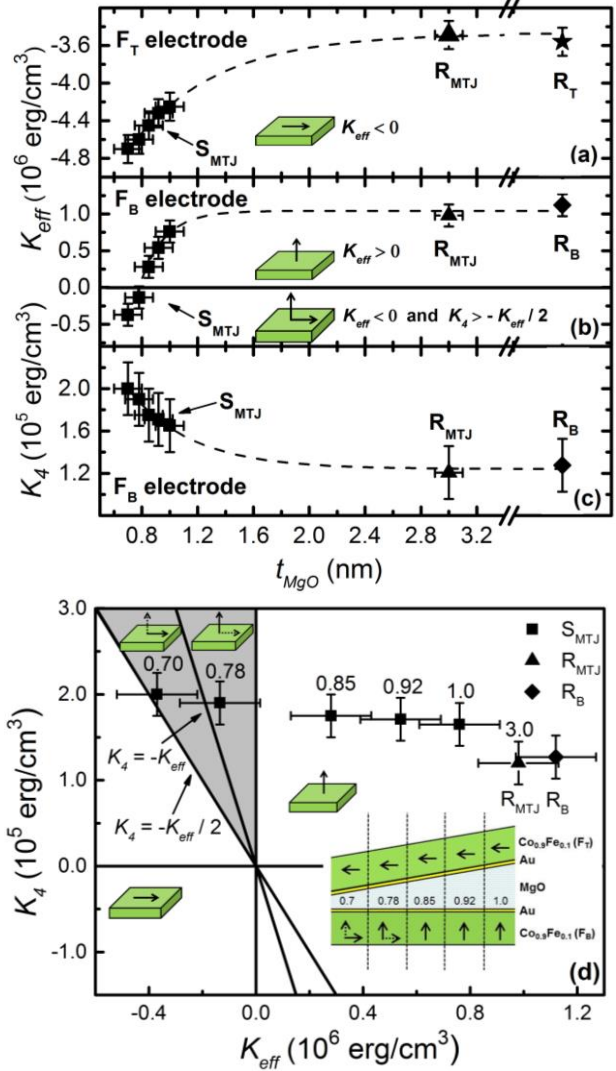

Fig. 6. Symbols: The effective uniaxial $K_{eff}$ (a), (b) and second-order $K_4$ (c) anisotropy constants as a function of the MgO film thickness obtained for the $S_{MTJ}$ and $R_{MTJ}$ structures; symbols after the axis break represent the data obtained for reference $R_B$ and $R_T$ samples. The dashed lines serve as a guide to the eye; (d) Stability regions for the easy axis of magnetization. Symbols: the second-order $K_4$ versus the effective uniaxial $K_{eff}$ anisotropy constants of the $F_B$ electrode of the samples $S_{MTJ}$, $R_{MTJ}$, and $R_B$. The numbers above the data points indicate the MgO layer thickness in nm. The inset illustrates the evolution of orientation of the easy axis of magnetization in the bottom $F_B$ electrode. Dotted arrows in the insets indicate metastable directions of magnetization. The scale of layer thicknesses and sample length are not preserved.

## 5. Discussion

According to the theoretical and experimental analysis of the magnetic stability regions, [23,24] the SRT process can proceed over three different paths, *i.e.* through: the point with zero anisotropy, a coexistence of easy-axis and easy-plane anisotropy or a region with canted magnetization. Each of these stability regions can be described by the set of anisotropy parameters. Namely, for positive $K_{eff}$ an out-of-plane easy direction of magnetization will appear when $K_4 > - K_{eff}/2$, while for $K_4 < - K_{eff}/2$ the canted magnetization will be realized. For negative $K_{eff}$ and $K_4 < - K_{eff}/2$ there is an easy plane of magnetization, while for $K_4 > - K_{eff}/2$ the coexistence of easy-plane and easy-axis of magnetization is observed. As it is demonstrated in the $K_4$ versus $K_{eff}$ diagram presented in the Fig. 6(d) the magnetization of the $F_B$ electrode of the $S_{MTJ}$ samples is out-of-plane down to $t_{MgO}$ of about 0.8 nm. When $t_{MgO}$ decreases below 0.8 nm, the anisotropy constants satisfy the conditions for the coexistence of the two easy directions of magnetization, one along the z-axis (stable) and the other one in the sample plane (metastable). For the further reduction of $t_{MgO}$, two easy directions are still observed, but now the direction along the z-axis is metastable, and the direction in the sample plane is a stable one.

The observed MA evolution for the $t_{MgO} < 0.8$ nm can be attributed to a presence of an interlayer exchange coupling (IEC) between $Co_{0.9}Fe_{0.1}$ layers through the thin MgO film. This hypothesis is in agreement with the theoretical predictions, [25,26] and experimental works, [27,28] wherein it was shown that in the FM/I/FM systems one should expect the IEC when the thickness of the spacer is lower than 0.8 nm. Moreover, it was shown that oxygen vacancies, as well as interfacial oxygen, cause the IEC to be antiferromagnetic for MgO thickness below 0.8 nm. [29,30] Indeed, accomplished FMR experiments show the continuous falloff of the intensity of the resonance line of the $F_B$ layer as the thickness of MgO is reduced below 1 nm. This observation, in turn, suggests that the magnetic moments on the surface are progressively affected by the interlayer interactions and as a result, the surface anisotropy energy and, consequently, the MAE, changes. When the thickness of the insulator is further reduced, the contribution of an in-plane anisotropy energy components increases and the coexistence of vertical and in-plane easy directions shows up.

## 6. Summary

The interaction between the two magnetic layers, separated by a thin insulating MgO layer, in the Mo/Au/$Co_{0.9}Fe_{0.1}$/Au/MgO($t_{MgO}$)/Au/$Co_{0.9}Fe_{0.1}$/Au structure, and the impact of this effect on the magnetic anisotropy of the $Co_{0.9}Fe_{0.1}$ layers has been presented. The results of the FMR and magnetometry investigations have demonstrated that the magnetic anisotropy of the bottom $Co_{0.9}Fe_{0.1}$ layer depends on the MgO spacer thickness. It was shown, that the magnetization is out-of-plane down to $t_{MgO}$ of about 0.8 nm, and below this thickness, the coexistence of perpendicular and in-plane easy direction of magnetization is observed. The corresponding zeroing of the effective uniaxial anisotropy constant allows us to establish the magnitude of the second-order uniaxial component to the magnetic anisotropy of the bottom layer. It amounts to $(1.9 \pm 0.5) \times 10^5$ erg/cm$^3$, meaning that in this range of $t_{MgO}$ the magnetic anisotropy of the bottom layer acquires a fourfold character. An interlayer exchange coupling (IEC) between $Co_{0.9}Fe_{0.1}$ layers through the thin MgO film was proposed as a possible mechanism of the origin of the observed spin reorientation transition. Presented results are giving new insight into spin reorientation phenomena in the MTJ structures which is of great technological interest.

## Acknowledgments

This work is partially supported by the National Science Centre in Poland under the project no. 2014/13/B/ST5/01834.